\documentclass[prl, twocolumn, nofootinbib, superscriptaddress]{revtex4-1}    

\usepackage{amssymb}
\usepackage{amsmath}
\usepackage{amsthm}
\usepackage{graphicx}
\usepackage[dvips]{epsfig}
\usepackage[usenames,dvipsnames]{color}
\usepackage{hyperref}
\usepackage{mathbbol}

\hypersetup{
   colorlinks=true,         
   linkcolor=blue,          
   citecolor=red,        
   urlcolor=Violet            
}

\newcommand{\be}{\begin{equation}}
\newcommand{\ee}{\end{equation}}

\newcommand{\fLie}{{\mathbb{L}}}
\newcommand{\fI}{{\mathbb{i}}}
\newcommand{\F}{{\Phi}}
\renewcommand{\d}{{\mathrm{d}}}
\newcommand{\D}{{\mathrm{D}}}

\newcommand{\G}{{\mathcal{G}}}
\newcommand{\Ad}{{\mathrm{Ad}}}
\newcommand{\pp}{{\partial}}
\newcommand{\fvf}{{\mathbb{X}}}
\newcommand{\lvf}{{\xi}}
\newcommand{\fv}{{\mathbb{a}}}
\newcommand{\FYM}{{\Phi_\text{YM}}}

\newcommand{\FM}{{\Phi_\text{Max}}}
\newcommand{\FQED}{{\Phi_\text{QED}}}

\newcommand{\PPexp}{{\mathbb{P}\hspace{-1pt}\exp}}
\newcommand{\fG}{{\mathrm{Lie}(\G)}}
\newcommand{\fX}{{\mathfrak{X}}}

\renewcommand{\bar}{\overline}
\newcommand{\dd}{{\mathbb{d}}}
\newcommand{\la}{\langle}
\newcommand{\ra}{\rangle}
\renewcommand{\hat}{\widehat}

\newcommand{\cint}{{\int\kern-.87em{<}}}
\newcommand{\sint}{{\int\kern-.75em{\sim}}}
\newcommand{\fint}{{\int\kern-.57em{\int}}}

\renewcommand{\#}{\sharp}

\let\oldmarginpar\marginpar
\renewcommand\marginpar[1]{\oldmarginpar{\color{red}\raggedright\footnotesize #1}}

\begin{document}

\title{A Unified Geometric Framework for Boundary Charges and Particle Dressings}
\author{Henrique Gomes} 
\author{Aldo Riello}
\affiliation{Perimeter Institute for Theoretical Physics, Waterloo, Ontario N2L 2Y5, Canada}

\begin{abstract}
We provide a unified geometrical origin for both boundary charges and particle dressings, with a focus on  electrodynamics. The method is furthermore generalizable to QCD and gravity, and can be extended to the  \lq{}non-perturbative\rq{} domain. 
\end{abstract}

\maketitle


Purely geometrical tools  often give valuable insight into physical questions. Examples abound: 
from the role of Riemannian geometry in general relativity, to the use of the Atiyah--Singer index theorem in studying anomalies in quantum field theories, with many examples in-between, before, and more recently. 

We propose here that another geometrical tool has wide applications in gauge theories: the \textit{field-space connection-form}, here called $\varpi$  (pronounced `var-PIE') \cite{GomesRiello}. Much like its finite-dimensional cousin $A_\mu$, the geometric role of $\varpi$ is to implement gauge-covariance. Differently from its finite-dimensional cousin, $\varpi$ implements covariance in the \textit{field-space} of gauge theories.

Beyond being a mere mathematical curiosity, this tool is surprisingly powerful, both in the study of boundary charges  and in the characterization of the dressings of charged particles in gauge theories. 

These two topics---boundary charges and dressings---were related by Bagan et al \cite{LavelleQCD}, upon defining constituent quarks as color-charged gauge-invariant entities. Such entities were built out of a Lagrangian quark which was then dressed by a cloud of gluons; a construction analogous to the Dirac dressing of electrons \cite{Dirac,LavelleDressing}.

More recently, charges and dressings have prominently come together in a series of works where enlarged asymptotic symmetry groups and the associated conserved charges {were} related to memory effects and soft-photon dressings (see \cite{StromingerLectures} and refs therein).

On  possibly related developments, new boundary degrees of freedom have been deemed necessary and introduced both to reinstate gauge-invariance in the presence of boundaries \cite{Waidia,DonnellyFreidel}  and/or to account for the correct entanglement entropy of gauge theories in finitely bounded regions (such as black hole spacetimes)  \cite{ReggeTeitelboim,Carlip2005,DonnellyWall}.

As has become clear in the study of asymptotic conditions on spacetime, there is ambiguity in how we parse \lq{}pure gauge\rq{} transformations  from the global symmetries. Prominent examples are the subtle choices of fall-off (and parity) conditions in \cite{ReggeTeitelboim, BrownHenneaux}. These choices are consequential: they translate to different asymptotic charges and associated algebras \cite{BarnichBrandt}.  Their ambiguity represents different answers to the question `which are the gauge degrees of freedom---the ones to be arbitrarily fixed---and which are not?' Or in other words, in the presence of (asymptotic) boundaries, how do we tell when a gauge-fixing has gone too far? 

The field-space connection-form $\varpi$ \cite{GomesRiello} can provide a common source of  explanation and an organizing principle to many aspects of the above-mentioned questions: it reconciles gauge degrees of freedom and boundaries, rendering the  introduction of new boundary degrees of freedom superfluous. Maintaining covariance means $\varpi$ always keeps track of all the degrees of freedom---including the possibly gauge ones. Nonetheless, \lq{}true\rq{} gauge transformations only give rise to vanishing charges, while global charges still emerge from the formalism. Given the aforementioned choices and ambiguities, this is a significant advance. 

Moving forward, $\varpi$ has a straightforward relation to \lq{}dressings\rq{} \cite{Dirac,LavelleDressing, StromingerLectures}. More importantly, it also provides a clear geometric path for obtaining dressings  in non-Abelian theories, even in the non-perturbative setting---an area other notions of dressings which rely on gauge-fixings \cite{LavelleDressing} cannot reach due to the so-called Gribov prolem \cite{Gribov,Singer}.

\paragraph*{Summary of results.}%
After introducing concepts and notation for dealing efficiently with the geometry of field-space, we will show how a simple choice of $\varpi$, naturally related to the dynamics of a gauge theory, readily provides a notion of dressing. This is found to coincide with the Dirac dressing in the context of 3+1 electrodynamics. 

We then show that using $\varpi$-covariant symplectic geometry produces vanishing charges for `pure gauge' transformations, even in the presence of boundaries and when the gauge parameters are field-dependent. 
Physically, this happens because our formalism automatically includes those contributions to the  charges which can be attributed to dressings. 
Moreover, we will show that using Dirac-like dressings, the global conserved charges---such as the total electric charge---are naturally picked out as the only physical ones. 

Further results and explicit examples of field-space connections will appear in a forthcoming publication \cite{GomesHopfRiello}.


\section{The field-space connection-form}

\paragraph*{Field space preliminaries.}

Consider the space of fields $\varphi^I$ defined on an $n$-dimensional manifold $M$, $\F=\{\varphi^I\}$.
In this notation, $\varphi^I$ stands for a whole field configuration $\{\varphi^I(x)\}_{x\in M}$, where $I$ is a super-index labeling both the field's type and its various components. In the following, a `double-struck' typeface---like in $ \dd$, $\fLie$, $\fvf$, etc.---will be consistently used for field-space entities.

On $\F$, introduce the deRahm differential $\dd$ \cite{Witten}; it should be thought as the analogue, on $\F$, of the spacetime differential $\d$.
A basis of $\Lambda^1(\F)$ is hence given by $\Big(\dd \varphi^I(x) \Big)$.
On a functional $f:\F\to \mathbb{R}$ (reals), $\dd$ acts as: 
\begin{align}
\dd f = \sum_{I}\int_M \d^n x \left( \frac{\delta f}{\delta  \varphi^{I}(x)} \dd \varphi^{I}(x)\right)  =:\int \frac{\dd f}{\dd \varphi^I} \dd \varphi^I,
\label{dd}
\end{align}
where $\delta/\delta \varphi$ denotes as usual a functional derivative, and the last identity introduces a more homogeneous short-handed notation.
Higher dimensional (functional) forms are defined by the above formula and antisymmetrization. In particular $ \dd^2 = 0$ (wedge products are left understood).

Functional (spacetime-local) vector fields on $\F$ are denoted $\fvf \in \mathfrak{X}^1(\F)$. In components, they read
\be\label{vectors}
\fvf =  \sum_I \int_M \d^n x \left( \fvf^I(\varphi(x)) \frac{\delta}{\delta \varphi^I(x)}\right) =: \int \fvf^I\frac{\dd}{\dd \varphi^I} \,,
\ee
where the introduced notation follows that of \eqref{dd}.
Contraction (`inclusion') of a vector field with a differential form in $\F$ is denoted with $\fI$, and defined by
\be
\fI_\fvf \dd \varphi^I = \fvf^I
\ee
and the usual rules of linearity and antisymmetrization. 

Finally, we introduce the functional Lie derivative along $\fvf$ of a generic functional form through the Cartan formula
\be
\fLie_\fvf = \fI_\fvf\dd + \dd \fI_\fvf.
\ee

\paragraph*{Gauge theories and the connection-form.}

The field space of Yang--Mills theory coupled to matter is given by a gauge potential $A$ and (spinorial) matter fields $\psi$, $\FYM = \{ \varphi^I = (A,\psi)  \}$, where we suppressed spacetime, spinorial, Lie algebra and representation indices. 

The group of gauge transformations is taken pointwise in the space(time) manifold\footnote{ Here $M$ may be either a spatial or spacetime manifold. We will specialize to a spatial one later.} $M$, i.e. $\G = \{g(\cdot):M\to G\},$ and elements $g(\cdot)\in\G$ act on the fields infinitesimally, with ${\lvf} \in \fG$, as $A \mapsto A + \delta_\lvf A$ and $\psi \mapsto \psi +\delta_{\lvf} \psi$, where 
\be
\delta_\lvf A:=\D_A {\lvf} = \d {\lvf} + [A,{\lvf}], 
\qquad
\delta_{\lvf}\psi:= - {\lvf} \psi,
\label{eq_deltaY}
\ee
and $[\cdot,\cdot]$ is the Lie bracket of $\mathfrak{g}:=\mathrm{Lie}(G)$. 
This defines a lift from the Lie algebra of the gauge group, $\fG$, into field-space vector fields $\fX^1(\FYM)$ \cite{KobayashiNomizu}
\be
\fG \to \fX^1(\FYM) , \quad {\lvf} \mapsto {\lvf}^\#(\varphi) :=\int \delta_{\lvf} \varphi^I \frac{\dd}{\dd \varphi^I}.
\ee
The map $\cdot^\#$ has a trivial kernel if $\psi\neq0$.
The vector fields $\xi^\#$ are canonically defined. 
Their flows generate gauge orbits in $\FYM$, which can be interpreted as the fibers of an infinite dimensional principal fiber bundle $\G \hookrightarrow \FYM \xrightarrow{\pi}[\F_\text{YM}]$, where $[\F_\text{YM}]:= \FYM/\G$ is the reduced space of physical field configurations.   This picture emerges in case the action of the group is free, which is not the case for most gauge theories. When the group action on $\F$ has fixed points, {i.e. there are configurations which are left invariant by  `Killing' gauge transformations,}  $\F/\G$ is not a manifold but a stratified manifold \cite{Fischer}. The strata  will turn out to be related to the conserved global charges. 

General vector fields $\fvf$ which are tangent to  gauge orbits in $\FYM$ will be called `vertical' and their span at a  $\varphi\in\FYM$ defines a vertical subspace of the tangent space. In symbols, $ \mathrm T_\varphi \FYM \supset {V}_\varphi = \mathrm{Span}\{{\lvf}^\#, {\lvf}\in\fG\}$.
Vertical fields represent infinitesimal, possibly field-dependent  gauge transformations (for $\xi:\FYM\rightarrow\fG$).

Crucially, there is no canonical transversal complement to the vertical subspaces, $\mathrm T\FYM\simeq {V}\oplus {H}$, where ${H}$ is (a choice of) `horizontal' subspace. 
Locally, a choice of $H$ corresponds to the choice of a vertical projector, $\hat V_\varphi:\mathrm T_\varphi\FYM\to V_\varphi$.
If  one then requires  the projector to be compatible with the  gauge-orbit structure of $\FYM$, one is led to introduce a $\fG$-valued functional 1-form $\varpi\in\Lambda^1(\FYM,\fG)$,  for which  $H := \{ \fvf \in \mathrm T\FYM \,|\, \fI_\fvf \varpi = 0\}$, and which satisfies the properties of a \textit{connection-form}:
\be
\;\;
\fI_{{\lvf}^\#} \varpi = {\lvf}\,,
\;\;\;\;
\fLie_{{\lvf}^\#} \varpi =  [\varpi, {\lvf}] + \dd {\lvf}.
\label{conditions}
\ee
The last term of the last formula accounts for field-dependent gauge transformations \cite{GomesRiello}.  

Concretely, this construction replaces the ordinary exterior derivative in field space, $\dd$, with a {\it covariant} or, more precisely, `horizontal' version. 
{For field-space scalars, e.g. all the $\varphi^I$, this is given by
\be
\dd_H=\dd - \delta_{\varpi} 
\ee
($\varpi$ being valued in $\fG$, $\delta_\varpi$ is defined as in \eqref{eq_deltaY}),
while in the case of the field-space connection, its horizontal variation defines the field-space curvature
\be
\mathbb{F}:=\dd_H \varpi = \dd \varpi + \tfrac12 [\varpi,\varpi].
\ee
These definitions are in complete analogy with the finite dimensional principal fiber bundle picture of gauge theory, see e.g. \cite{KobayashiNomizu}.}

 In \cite{GomesRiello}, we argued that horizontal field variations, i.e. horizontal tangent vectors in $\F$, have the interpretation of `{\it physical} changes with respect to to the choice of $\varpi$'. In the following we will flesh this out.

\section{Electrodynamics}

The simplest notion of horizontality---and thus of $\varpi$---is given by orthogonality to $V$ with respect to a metric on field space.
Such a metric is required to be invariant along the gauge orbits  to ensure the covariance of $\varpi$ \cite{DeWitt, GomesRiem}.

We now specialize to a simple example.
Let us consider Maxwell theory in a 3+1 decomposition on $M=\Sigma\times\mathbb R$, and let us take field space to be the space of `instantaneous configurations', $\FM:=\{A_i(x)\}_{x\in\Sigma}$, {and $\G_\text{Max}:=\{g(\cdot):\Sigma\times\mathbb{R}\to \mathrm{U}(1)\}$.} Field histories,  $A_i(x,t)$, are curves in this space. Here, we will denote general vectors at $A$ by $\mathbb{a}$, i.e. $\mathbb{a}\equiv\int\mathbb{a}_i \frac{\dd}{\dd A_i}\in \mathrm{T}_{A}\FM$.

The last component of the electromagnetic potential, that is,  $A_0$, is a Lagrange multiplier that can be freely fixed to any function $\lambda(x,t)$ along a curve $A_i(x,t)$. To ensure covariance with respect to history-dependent gauge transformations, we add an extra $\varpi$-dependent term, $A_0 = \lambda + \fI_{\mathbb t}\varpi$. Here, ${\mathbb t} = \int {\dot A}_i \frac{\dd}{\dd A_i}\in \mathrm{T}_{A}\FM$ is the \lq{}evolution vector\rq{} along a curve $A(x,t)\subset\FM$, the dot meaning derivation with respect to $t$ (the evolution need not be on-shell). Note, $\fI_{\mathbb t}\varpi$ gauge-transforms in the  manner expected of $A_0$.

Define on $\FM$ the constant DeWitt (super-)metric $\la \cdot, \cdot \ra$ to be given by the ultralocal contraction of two tangent vectors $\fv, \fv'\in \mathrm{T}_{A}\FM$ through the inverse metric $g^{ij}$ of $\Sigma$, 
\be
\la \fv, \fv' \ra_A = \int_\Sigma \d^3 x  \,\sqrt{g} g^{ij} \mathbb{a}_i(x)\mathbb{a}'_j(x)
\label{metric}
\ee
(if $\Sigma$ is non-compact, appropriate fall-off conditions at spatial infinity are presupposed for normalizability.) 

As is easy to see, this is the field-space metric which contracts $\dot A_i$ (the components of $\mathbb{t}$) in the kinetic term of the Lagrangian.  In this sense, this metric is compatible with the phase space structure of the theory, and therefore constitutes a dynamically preferred choice.%
\footnote{The knowledge of the kinematic terms of the Lagrangian together with the demand of minimal coupling (gauge structure) is enough to reconstruct the full dynamics of gauge theories coupled to matter.}
This is also the reason why we will sometimes refer to the DeWitt metric as a `kinematical metric'.

Vertical vectors at $A_i$ are spanned by $\mathbb{a}_i=\delta_{\lvf} A_i=-i\, \pp_i {\lvf}$, where\footnote{We adopt the anti-hermitian convention for $\frak g$ when $G$ is unitary, but we still keep the Maxwell field real---so that it is valued in $-i \frak g$. For notational simplicity, we fix the electron charge to $+1$.} ${\lvf}\in \mathrm{Lie}(\mathcal G_\text{Max}) \cong  i\,C^\infty(\Sigma)$. 
Notice that, in the present Abelian case, the $\delta_{\lvf} A_i$ are field-independent.  
It is easy to find the orthogonal (horizontal) complement of the vertical vectors $\int \delta_{\lvf}A\frac{\dd}{\dd A}$ from \eqref{metric}. 
From this, using that the horizontal projection of a generic ${\fv}$ is ${\fv} - (\fI_{\fv}\varpi)^\#$, one derives that at $A\in\FM$ the resulting $\varpi$ must satisfy a Laplace equation with Neumann boundary conditions:
\be\label{YM_varpi_boundaries}
\nabla^2 \varpi = i\, \mathrm{div}(\dd A),
\quad
n^j\nabla_j \varpi|_{\pp \Sigma} = i \,n\cdot\dd A|_{\pp \Sigma},
\ee
where $n^i$ is the (spacelike) normal to $\partial\Sigma$, if this is not empty (see \cite{GomesHopfRiello} for details). 
The unique solution to this equation is
\be\label{varpi_QED}
\varpi(x)=i \Big(\nabla^{-2}\mathrm{div}(\dd A)\Big)(x)=  i\int_\Sigma \frac{\d^3 y}{4\pi} \frac{\pp^i\dd A_i(y)}{|x-y|},
\ee
where for definiteness we fixed $\Sigma\cong\mathbb R^3$ with fast decaying boundary conditions. This expression for $\varpi$ satisfies \eqref{conditions}.

We name a connection form derived through the above algorithm a {\it DeWitt connection}.
In the case of Maxwell theory, where photons are uncharged and the $\delta_{\lvf} A_i$ are field-independent, the DeWitt connection is also field-independent. We shall come back to this point when we  discuss the generalization to non-Abelian theories.
Since the gauge transformation for the vector potential $A_i$ involves derivatives of ${\lvf}$, the resulting DeWitt connection turns out to be non-local.

Note that if we add (charged) matter fields to the pure Maxwell theory, obtaining $\FQED=\{(A,\psi)\}$, the above would still be a valid connection form on the {\it full} $\FQED$.

\paragraph*{Remark.} A more covariant treatment, which uses a space of histories for $A_\mu$ rather than a space of configurations for $A_i$ is possible in principle. Nevertheless, it requires choices of Green functions and introduces time non-localities in the construction of the connection. We leave the investigation of these aspects for future work.

\section{DRESSINGS}

For field-space curves $A_i(x,t)$, in analogy with a Wilson line, we define through a path-ordered exponential the field-dependent field-space \lq{}parallel-transport\rq{}
\be
h[A] = \PPexp \int_{A^\star \leftarrow A} \varpi\,,
\label{h}
\ee
where $(A^\star\leftarrow A)=\{A(t)\}$ is a field-space path linking the configuration $A$ to the initial configuration $A^\star=A(0)$, the arrow indicating the direction of path-ordering. 

Under a field-dependent gauge transformation $g[A](\cdot)\in\G$, \eqref{h} transforms  at every point $x\in\Sigma$ as $h[A]\mapsto g[A]^{-1}h[A]g[A^\star]$, as follows from \eqref{conditions}. 
We consider the initial configuration to be a fully fixed reference configuration, so that $g_\star\equiv g[A^\star]=\mathrm{Id}$.

Now, define the \textit{dressed} matter and gauge fields by
\be
\hat \psi = h^{-1} \psi
\quad\text{and}\quad
\hat A = A^{h}= A + h^{-1}\d h.
\label{dressed}
\ee
Under the action of $g[A]$, the dressed fields  transform into
  $ (\,\hat A + g_\star^{-1}\d g_\star \;,\; g_\star^{-1}\hat\psi \,)\stackrel{g_\star=\mathrm{Id}}{=}\; (\hat A,\hat\psi).$
 In other words, under gauge transformations which leave $A^\star$ fixed, the corresponding dressed fields are fully gauge invariant.

In the case of Maxwell theory, the DeWitt connection given in \eqref{varpi_QED} is independent of $A$, and consequently $h[A]$ is path-independent in field space. 
If (for mere convenience) we choose $A^\star$ to be in the gauge $\pp^iA_i^\star=0$, then
\be
h[A] = \exp\Big(i\,  \nabla^{-2}\mathrm{div} A\Big)
\ee 
is readily recognized to be the Dirac dressing.
Henceforth, we will call the field-space connection form of \eqref{varpi_QED}, the {\it (kinematical) Dirac--DeWitt connection}.

\paragraph{Remark.} To make contact with the Faddeev--Kulish dressing \cite{FK,GomezPanchenko}---the one relevant for soft-charges \cite{StromingerLectures}---we briefly note that in Lorentz gauge, which in momentum space reads $p^\mu \tilde A_\mu=0$, we obtain $h[A]= \exp\left(\frac{i}{2\pi} \int \frac{\d^3 p}{2E_p}\frac{p^i \tilde A_i}{p^jp_j}\right)$ which coincides on-shell ($p^ip_i=E_p^2$)  with the Faddeev--Kulish dressing in the rest-frame of the electron. While to make explicit contact with \cite{LavelleDressing,LavelleShort}, we note that their two fundamental demands of a static dressing correspond, irrespectively of gauge-fixings, to the first condition of \eqref{conditions}, and to setting $\lambda(x,t)=0$ in the definition $A_0 = \lambda + \fI_{\mathbb t}\varpi$.

\section{Local and global charges}

In the presence of finite boundaries,\footnote{These should be understood as boundaries of a subregion of $\Sigma$.} gauge-invariance can pose a challenge (e.g. \cite{ReggeTeitelboim, Waidia, BrownHenneaux, LeeWald, Wald, WaldZoupas, Seraj, StromingerLectures, DonnellyFreidel}), especially if one has in the formalism field-dependent gauge transformations, implied by dressings of all sorts. In this section, we will show that even in the presence of boundaries, using $\dd_H$ as opposed to $\dd$ in the spacetime-covariant symplectic approach \cite{GomesRiello}, allows us to gain complete \text{local} gauge-invariance, while retaining---when using the Dirac--DeWitt connection---solely the physical  conserved charges.
We will also show that, in light of the previous section, $\dd_H$-symplectic geometry corresponds to the symplectic geometry of the dressed fields.

We start  by recalling the construction of the charges in the symplectic language.
Whenever the Lagrangian {\it density} $\mathcal{L}(\varphi)\d^4 x$ is invariant under gauge transformations, as in Yang--Mills,\footnote{In General Relativity there are subtleties with boundary terms \cite{LeeWald,WaldZoupas}.}
\be
0 = \fLie_{{\lvf}^\#} \mathcal{L}\d^4 x = \mathrm{EL}_I\delta_{\lvf}\varphi^I \d^4 x + \d\theta(\varphi,\delta_{\lvf}\varphi) ,
\ee
here $\mathrm{EL}_I(\varphi)$ are the Euler-Lagrange equations for $\varphi^I$, and $\theta=\Pi_I\dd\varphi^I\in \Lambda^1(\F)\otimes \Lambda^{3}(M)$ is standard notation for the (pre)symplectic current (we use densitized momenta $\Pi_I$). One can define the (on-shell) conserved Noether current density $j_{\lvf}$ as (e.g. \cite{Wald}) 
\be
j_{\lvf} := \fI_{{\lvf}^\#} \theta \equiv \theta(\varphi,\delta_{\lvf}\varphi).
\ee
In particular, the extra invariance $\fLie_{{\lvf}^\#}\theta=0$ implies, via the Cartan formula, the Hamiltonian flow equation $\fI_{{\lvf}^\#}\Omega=-\dd j_{\lvf} $ thus indicating a symmetry generator ($\Omega=\dd\theta$ is the (pre)symplectic two-form).

In Yang--Mills theories it is easy to show that  the Noether current density is  exact,  $j_{\lvf}\approx \d(E\lvf)$, when on-shell of the Gauss constraint (a condition we signal with $\approx$). Hence the associated charge is a pure boundary quantity. This is why one talks always about `boundary charges'. 

Now, $\fLie_{{\lvf}^\#}\theta=0$ holds in Yang--Mills theories only for {\it field-independent} gauge transformations, i.e. only if $\dd {\lvf}=0$.
This led us to introduce the horizontal symplectic current \cite{GomesRiello}, 
\be
\theta_H :=\Pi_I\dd_H\varphi^I = \theta + \Pi_I\delta_{\varpi}\varphi^I,
\qquad
\fLie_{{\lvf}^\#}\theta_H \equiv 0.
\ee
The last equality is easily checked in Yang--Mills theory.
It then follows that $\Omega_H:=\dd_H\theta_H=\dd\theta_H$ is $\dd$-exact---which makes it a viable presymplectic form---and
\be
j^H_{\lvf} := \fI_{{\lvf}^\#}\theta_H =  0. 
\label{hor_charge_gauge}
\ee
This formula is valid locally on $M$, at the density level.
The message it conveys is that gauge transformations carry {\it no physical charge}  with respect to this particular decomposition of vertical-horizontal (or gauge-physical) degrees of freedom. However, there is still room for  conserved global charges. 

Before addressing global charges, one remark is in order.
The symplectic potential as derived from $\mathcal L$ is defined up to a $\d$-exact term. 
In Yang-Mills theory,  it is precisely such a boundary term that distinguishes $\theta$  from $\theta_H$  \cite{GomesRiello, DonnellyFreidel}, 
\be
\theta_H\approx \theta + \d(E\varpi),
\ee 
and similarly, apart from boundary terms, $(\Omega_H)_{|\text{bulk}}\approx \Omega_{|\text{bulk}}$.
{As customary in gauge theories and general relativity, pure boundary contributions can be highly non-local: although the image is restricted to the boundary, their domain depend on the fields throughout the whole manifold---this is the case for the Dirac--DeWitt connection \eqref{varpi_QED}, but need not be for other choices \cite{GomesHopfRiello}.}

\paragraph*{Global charges.}

So far we have implicitly assumed that  $\varpi$ provides   a 1-1 relation between $\fG$ and $V_\varphi$. 
 In practice, this is not always the case,  even if the operator $\cdot^\#:\fG\rightarrow V_\varphi$ is pointwise in $\Phi$ an isomorphism (we assume $\psi\neq0$); there may exist particular ${\lvf}_o\in \fG$ for which $\fI _{{\lvf}_o^\#}\varpi=0$.
Such ${\lvf}_o^\#$'s---if they exist---are thus horizontal with respect to $\varpi$. Therefore, according to our identification `horizontal'$\sim$ `physical', the transformations  corresponding to ${\lvf}_o$'s play the role of actual symmetries, not of `unphysical' gauge transformations.

For the specific example of the Dirac--DeWitt connection on $\FQED$, from \eqref{varpi_QED} one sees that $\nabla^2 {\lvf}_o=0$ is a sufficient condition for $\fI _{{\lvf}_o^\#}\varpi=0$.
From \eqref{YM_varpi_boundaries} one infers that this condition is also necessary, and moreover,   that ${\lvf}_o$ has to satisfy vanishing Neumann boundary conditions. Hence, we conclude that for the Dirac--DeWitt connection, the only ${\lvf}_o$'s satisfying $\fI_{{\lvf}_o^\#}\varpi =0$ are constant ${\lvf}_o$'s (we assume $\Sigma$ has trivial cohomology). 

The physical relevance of the symmetry transformations ${\lvf}_o=cnst.$ is confirmed by the non-vanishing of the corresponding horizontal Noether current 
\begin{eqnarray}
j^H_{{\lvf}_o} &=&  \fI_{{\lvf}_o^\#} \Big[ E \wedge \dd_H A + \Big(\bar\psi \gamma_\mu \dd_H \psi \Big)\ast\d x^\mu \Big]  \notag\\
& =&   - {\lvf}_o \bar \psi \gamma_\mu \psi \ast \d x^\mu = - {\lvf}_o j_{\mathrm e},
\end{eqnarray}
where $\bar\psi := \psi^\dagger \gamma^0$ and $j_{\mathrm e}$ is the electron current density.

Thus we see that the Dirac--DeWitt connection automatically picks out global gauge transformations in electromagnetism as being physically distinct from local ones. 
 This is in contrast with those formalisms involving new boundary degrees of freedom, which tend to provide infinitely many boundary charges, one for each multipole moment of $\xi_{|\text{bdry}}$ \cite{DonnellyFreidel}.

The non-Abelian and gravitational analogues of electromagnetism's global gauge transformations are `Killing' gauge transformations and diffeomorphisms \cite{GomesHopfRiello}. Due to non-linearities, their very existence crucially depends on the properties of the field configuration, such as e.g. a metric $g_{\mu\nu}$ possessing Killing vector fields or a gauge potential $A^a_\mu$ being reducible.
A similar result was discussed by DeWitt \cite{DeWitt}. There, Killing transformations are picked out as the only actual symmetries by the non-linearities of the theory.

\paragraph*{The horizontal symplectic potential equals the dressed symplectic potential.}
 Finally, we show that gauge charges vanish thanks to the contribution of the dressings---this follows from equivalence between the use of a  horizontal symplectic potential  and that of \lq{}dressed\rq{} fields \eqref{dressed}. It is enough to show that $ \dd_H A = \Ad_{h}\,\dd \hat A $ and $\dd_H \psi =  h \dd\hat \psi$ where $\Ad$ is the adjoint action of the group (the notation encompasses the non-Abelian case). 
 The definition of $h$ via a path-ordered exponential \eqref{h} suffices to show that\footnote{$[\dd, \d]=0$ implies $\dd (h^{-1}\d h) = \d (h^{-1}\dd h) + [ h^{-1} \dd h, h^{-1} \d h]$.} $\dd h h^{-1} = \varpi$, which implies the result.
This result can be summarized as
\be
\theta( \hat \varphi, \dd \hat \varphi) = \theta_H(\varphi,\dd \varphi).
\ee

\section{Outlook}

As anticipated, \lq{}field-space covariance\rq{} provides for the first time a unified geometrical origin to both boundary charges and particle dressings in electrodynamics. This advance will be important in two main areas (and their intersection): the study of boundary properties in gauge theories, and non-perturbative treatments of non-abelian gauge theories.

\paragraph{Boundaries in gauge theories.} After a complete gauge-fixing, one has decided once and for all what is physical and what is gauge, and  information related to the latter degrees of freedom is obscured if not lost. This becomes extremely relevant for gauge theories in the presence of boundaries, both asymptotic and not. For such boundaries may accidentally break or otherwise fix  certain gauge-symmetries, which one would like to preserve in the physics of the system \cite{ReggeTeitelboim, Waidia, BrownHenneaux, LeeWald, Wald, WaldZoupas, Seraj, StromingerLectures, DonnellyFreidel}. New degrees of freedom
are sometimes inserted  into the theory  to restore the sought-after invariance \cite{ReggeTeitelboim,Carlip2005,DonnellyWall,DonnellyFreidel}. The field-space connection form, $\varpi$, being covariant and not invariant, retains the information about gauge directions. Some of these directions can still manifest themselves as global charges, but charges associated to generic local gauge symmetries happily always vanish in  the field-space covariant setting. In other words, in the cases explored so far, $\varpi$ has defeated the purpose of new degrees of freedom at boundaries; nothing is lost with $\varpi$, so nothing needs to be restored. 

The introduction of $\varpi$ begs for applications in other scenarios where boundary degrees of freedom have been introduced, such as \cite{ReggeTeitelboim, HenneauxNew,Wieland}. In those contexts where there is still controversy, it could give a natural characterization of the true physical charges as opposed to the purely gauge ones.

\paragraph{Non-perturbative, non-Abelian gauge theories.}
Many of the specific properties seen here are particular to the Abelian case. For a non-Abelian gauge theory, $\varpi$ could still be defined through orthogonality with respect to the obvious generalization of the DeWitt kinematical metric for the gauge field, Eq. \eqref{metric}. In that case, $\varpi$ turns out to be field-dependent, and the field-space Wilson line becomes path-(or history)-dependent due to the presence of field-space curvature, $\mathbb{F}=\dd_H\varpi\neq 0$. Indeed, $\varpi$ {\it cannot} be everywhere flat, since that would imply there exists a global (horizontal) section, in contradiction to the findings of Gribov \cite{Gribov,Singer}.
Nonetheless,  a well-defined, non-perturbative dressing, which reduces to \eqref{varpi_QED} around $A^\star=0$ at lowest order in perturbation theory, {still exists}. 
In this context, what we have just described in the last sentence is essentially a geometrized version of the proposals of \cite{LavelleQCD}. Lastly, we note that a fully Lorentz-covariant $\varpi$ for QCD would lead to a dressing similar to the Gribov-Zwanziger kind \cite{Zwanziger, GribovZwanziger}.   Understanding the natural extension  to the non-perturbative regime in QCD which $\varpi$ provides, and its relation to Gribov-Zwanziger and confinement, is an interesting future direction. \\

\acknowledgments 
We would like to thank F. Hopfm\"uller for discussions and for reading an earlier version of this draft, and W. Wieland for comments. This research was supported  by Perimeter Institute for Theoretical Physics. Research at Perimeter Institute is supported by the Government of Canada through Industry Canada and by the Province of Ontario through the Ministry of Research and Innovation. 

~\\
~\\
~\\


\bibliographystyle{ieeetr}


\end{document}